%Paper: cond-mat/9404041
%From: kantor@cmt.harvard.edu (Yacov Kantor)
%Date: Thu, 14 Apr 94 10:18:13 EDT

%----------------Excess Charge in Polyampholytes-------------------%
%
%----------------Fonts and magnification----------------------------------%
\magnification=\magstep1
\font\titlefont=cmr10 scaled\magstep3  
%----------------General definitions--------------------------------------%
\newbox\leftpage \newdimen\fullhsize \newdimen\hstitle \newdimen\hsbody
\hoffset=0.0truein \voffset=0.20truein \hsbody=\hsize \hstitle=\hsize
\tolerance=1000\hfuzz=2pt \baselineskip=20pt plus 4pt minus 2pt
%-----------------Equation numbers----------------------------------------%
\global\newcount\meqno \global\meqno=1
\def\eqn#1#2{\xdef #1{(\the\meqno)}\global\advance\meqno by1 $$#2\eqno#1$$}
%-----------------Reference numbers---------------------------------------%
\global\newcount\refno \global\refno=1 \newwrite\rfile
\def\ref#1#2{{[\the\refno]}\nref#1{#2}}%
\def\nref#1#2{\xdef#1{[\the\refno]}%
\ifnum\refno=1\immediate\openout\rfile=refs.tmp\fi%
\immediate\write\rfile{\noexpand\item{[\the\refno]\ }#2}%
\global\advance\refno by1}

%-----------------Figure/Table captions-----------------------------------%
\def\figures{\centerline{{\bf Figure Captions}}\medskip\parindent=40pt}
\def\fig#1#2{\medskip\item{Fig.~#1:  }#2}

%-----------------Other definitions---------------------------------------
    \def\frac#1#2{{#1\over#2}}
   
%-------------------------------------------------------------------------%
%
%-----------------TITLE PAGE----------------------------------------------%
\nopagenumbers \hsize=\hsbody \pageno=0 ~ \vfill
\centerline{\titlefont Excess Charge in Polyampholytes}
\bigskip
\centerline{{\sl Yacov Kantor}($^*$)($^{**}$)($^{***}$) and
{\sl Mehran Kardar}($^*$)($^{****}$)}
\centerline{($^*$)Department of Physics, Massachusetts Institute of Technology,
Cambridge, MA 02139, U.S.A.}
\centerline{($^{**}$)Physics Department, Harvard University,
Cambridge MA 02138, U.S.A.}
\centerline{($^{***}$)School of Physics and Astronomy, Tel Aviv University,
Tel Aviv 69 978, Israel\footnote{\dag}{Permanent address}}
\centerline{($^{****}$)Department of Theoretical Physics, Oxford University,
Oxford OX1 3NP, U.K.}
\medskip

\vfill\centerline{\bf ABSTRACT}\nobreak\medskip\nobreak\par{
Polyampholytes (PAs) are polymers composed of  (quenched) random
sequences of positive and negatively charged monomeric groups. We show
that the radius of gyration, $R_g$, of a PA strongly depends on its overall
excess charge $Q$, and is very weakly influenced by other aspects
of the sequence. For $Q<Q_c\approx q_0 \sqrt{N}$, where $N$ is the
number of monomers of charge $\pm q_0$, the PA is
compact, while for $Q>Q_c$ it is stretched. Some aspects of this
behavior can be understood by analogy with the shape instability of
a charged drop where the Rayleigh charge plays the role of $Q_c$.
}
\vfill
\medskip
\line{PACS. 36.20.--r Macromolecules and polymer molecules\hfill}
\line{PACS. 35.20.Bm General molecular conformation and symmetry;
stereochemistry\hfill}
\line{64.60.--i General studies of phase transitions   \hfill}
\line{PACS. 41.20.Cv Electrostatics; Poisson and Laplace equations,
boundary--value  problems \hfill}
\vfill\eject\footline={\hss\tenrm\folio\hss}
%----------------Main text------------------------------------------------%
As the building blocks of living matter, long chain macromolecules have
been the subject of intense study. Whereas there is now a reasonably
firm basis for understanding the physical properties of homopolymers
considerably less is known about the heteropolymers of biological
significance. From a biologist's perspective, it is the specific
properties of a particular molecule that are of interest. After
all, the genetic information is coded by very specific sequences
of nucleic acids, which are in turn translated to the chain of
amino acids forming a protein\ref
\rgen{See, e.g., Creighton T.E., {\it Proteins: Their Structure and
Molecular Properties}, Freeman, San Francisco (1984).}.
In response to the van der Waals, hydrogen bonding, hydrophobic/hydrophilic,
and Coulomb interactions,
the protein folds into a `native' shape that is responsible for its activity.
Given the large number of monomers making up such chains,
finding the configuration of a particular molecule is a formidable task.
By contrast, a physicist's approach is to sacrifice the specificity,
in the hope of gleaming some general information from simplified
models\ref
\rstein{Stein D.L., Proc. Natl. Acad. Sci. USA {\bf 82}, 3670 (1985);
Bryngelson J.D. and Wolynes P.G., Proc. Natl. Acad. Sci. USA
{\bf 84}, 7524 (1987); Chan H.S. and Dill K.A., Physics Today, p. 24,
February 1993.}.
There are in fact several statistical descriptions of proteins
as random linear sequences of elements with a variety
of interactions that determine their final shape\ref
\rGOa{Garel T. and Orland H., Europhys. Lett. {\bf 6}, 307 (1988);
Shakhnovich E.I. and Gutin A.M., Europhys. Lett. {\bf 8}, 327 (1989);
Karplus M. and Shakhnovich E.I., in {\it Protein Folding}, ch.4,
p. 127, Freeman \& Co., New York (1992).}
\ref\rKK{Kantor Y. and Kardar M., Europhys. Lett. {\bf 14}, 421 (1991).}.

Statistical models only provide gross, general
descriptions of the heteropolymer. Perhaps the coarsest characteristic
of a molecule is some measure of its size, such as the radius of
gyration, $R_g$. For a polymer with charged groups, the overall size
is most likely controlled by the Coulomb interactions
which are the strongest and with the longest range. In this paper
we consider the typical size of {\it polyampholytes} (PAs): polymers
composed of positive and negatively charged groups\ref
\rPA{Tanford C., {\it Physical Chemistry of Macromolecules}, Wiley, NY, 1961.}.
Even determining the size of a simple model PA, formed from charges
$\pm q_0$ randomly dispersed along the chain, has proved controversial.
Appealing to a Debye--H\"uckel (DH) attraction
in analogy to electrolytes, Higgs and Joanny\ref
\rHJ{Higgs P.G. and Joanny J.-F., J. Chem. Phys. {\bf 94}, 1543 (1991).}\
conclude that a PA minimizes its (extensive) free energy by contracting
to a compact structure.
By contrast, in ref.\rKK, we used an analogy to the renormalization group
(RG) treatment of homogeneously charged polymers\ref
\rPVdG{Pfeuty P., Velasco R.M., and de Gennes P.G., J. Phys. (Paris) Lett.
{\bf 38}, L--5 (1977).}\
to conclude that the PA is stretched by the (non-extensive)
energy cost of typical charge fluctuations.
This contradiction was partially resolved\ref
\rKLKprl{Kantor Y., Li H.,
and Kardar M., Phys. Rev. Lett. {\bf 69}, 61 (1992);
Kantor Y., Kardar M., and Li H., Phys. Rev. {\bf E49},
1383 (1994).}\
by noting that the
DH theory requires exact neutrality of the electrolyte, while the
RG--inspired approach treats equally all possible sequences;
the typical charge of a random sequence,
$Q\approx \pm q_0 \sqrt{N}$, sufficing to modify the behavior of the PA.
Monte Carlo simulations\rKLKprl\ indeed confirm that PAs
with $Q=0$ compactify at low temperatures, while sampling all
random quenches with unrestricted $Q$ produces an average $R_g\propto N$.
Unfortunately, the broad nature of the distribution of sizes in the
latter ensemble provides little information about $R_g$ of a specific PA.

Recently, Yu {\it et al.}\ref
\rExT{Yu X.-H., Tanaka A., Tanaka K., and Tanaka T., J. Chem. Phys.
{\bf 97}, 7805 (1992);
Yu X.-H., Ph. D. thesis, MIT (1993).}\
performed a detailed investigation of the volume transition in gels
produced by crosslinking PAs.
The screening length in the experiments is quite big, enveloping
a large number of monomers $N$. By changing the pH of the solution,
it is possible to gradually modify the excess charge $Q$,
within the screening length. The neutral gel is naturally the
most compact. Surprisingly, the volume of the gel does
not change appreciably in an interval of $Q$ below a threshold
$Q_c\approx q_0 N^{1/2}$. Immediately beyond the threshold,
the volume of the gel increases by more than an order of magnitude.
These results motivated us to make a systematic examination
of the dependence of the $R_g$ of a PA on its excess charge.
Interestingly, we find that $Q$ is quite a good indicator of the
overall size of a PA, i.e. PAs with quite different sequences, but
the same excess charge, have a narrow distribution of $R_g$s.
The broad nature of the distribution of $R_g$s for unrestricted polymers
solely reflects the drastic variations of size with $Q$. In agreement
with experiments, we find that PAs are compact for small $Q$,
and  stretched when $Q$ exceeds a critical value of
$Q_c\approx q_0\sqrt{N}$. We provide a (partial) justification for
this transition by analogy to a charged drop: The drop is
stable at small $Q$, but disintegrates after suffering an instability
when $Q$ exceeds the Rayleigh charge $Q_R$.

The Monte Carlo (MC) procedure used in this work is identical to that
of ref.\rKLKprl.
We use self--avoiding chains on a cubic lattice of spacing $a$,
with electrostatic interaction $U_{ij}(r)=4q_iq_j/\sqrt{2a^2+r^2}$
for every pair ${\langle i,j\rangle}$ of charges at a distance $r$,
where $q_i=\pm q_0$ with signs specific to a particular quench.
We examined the dependence of polymer size on both temperature, $T$,
and excess charge, $Q$. Since we considered systems with quenched randomness,
the results were always averaged over 10 {\it different} quenches
for each $T$ and $Q$.
The MC time unit is defined as the interval over which $N$
attempts are made to move atoms. (Due to the long--range interaction,
CPU time per MC time unit grows as $N^2$.)
Each quenched configuration at a given temperature was equilibrated
for $250N^2$  MC time units, which for $N=64$ amounted to about 100
statistically independent configurations\ref
\rKKtopub{Kantor Y. and Kardar M., to be published (1994).}.
To collect all the data in Fig.1 we needed about two months of
CPU on a Silicon Graphics R4000 workstation.
Equilibration times were sufficiently long to insure that statistical
uncertainties of thermal averages were smaller than the differences
between different quenches.

Fig.1 depicts the temperature dependence of $R_g^2$ for 64--monomer
chains. The number near each curve indicates the charge, $Q/q_0$.
At very high temperatures the electrostatic
interactions are unimportant and the chains behave
as self--avoiding walks, with $R_g\propto N^\nu$ and $\nu=0.588$.
As $T$ is lowered, the effects of interactions are first observed
for the strongly charged polymers at $T_Q\approx Q^2/R_g\sim Q^2/N^{\nu}$
(measured in energy units).
This follows from the estimate of $Q^2/R_g$ for the electrostatic
energy of a typical configuration. However, for weakly
charged polymers, with $Q<q_0\sqrt{N}$, the main interaction
is between the non--homogeneities in the charge distribution.
The energy of a typical density fluctuation, $q(Q-q)/R_g
\approx -q_0^2 N/R_g$ for $q\approx q_0\sqrt{N}$ exceeds $Q^2/R_g$,
and such PAs start to deviate from their infinite--$T$ behavior
at $T_Q'\approx q_0^2 N/R_g\sim q_0^2N^{1-\nu}$.
These high temperature estimates suggest that on lowering $T$, the size
of PAs with large $Q$ increases (since the Coulomb energy of excess
charge is repulsive), while $R_g$ decreases for PAs with small $Q$
(since the charge fluctuation energy is attractive). This is confirmed
by a (high temperature) perturbative expansion of the radius in the
strength of the interaction as in ref.\rKLKprl. The first correction
is proportional to $\overline{q_iq_j}=(Q^2-q_0^2N)/N^2$ for $i\ne j$,
and the change in the behavior of $R_g$ occurs for $Q_c= q_0\sqrt{N}$.
(There is no a priori reason to expect higher order terms to
also vanish at $Q=Q_c$.)
Are the {\it averages} in Fig.1 a meaningful measure of the PA size?
Fig.2 depicts histograms of the distribution of $R_g^2$ at
$T=0.2q_0^2/a$ for several values of $Q$. Since thermal fluctuations
are small, this histogram represents the differences between
quenches. Note that the distributions are fairly narrow; their
widths not exceeding the distance between their averages. Thus the
location of a point in Fig.1 provides a good measure of $R_g^2(Q)$,
without need for any
additional reference to the details of the sequence. If this function
is known, the average of $R_g^2$ for {\it unrestricted} quenches
is simply obtained from,
$R_g^2({\rm random})=\int_0^\infty dQ R_g^2(Q) P(Q)$, where
$P(Q)\propto\exp[-Q^2/(2q_0^2N)]$ is the probability density of
an excess charge $Q$. In previous work\rKLKprl, we found that the
distribution of $R_g^2({\rm random})$ was very broad. Although the
sample of quenches was small, it included several completely collapsed
and  strongly stretched configurations.
The current results indicate that this follows simply from the strong
dependence of $R_g^2$ on $Q$, rather than a large scatter of $R_g^2$
among different quenches with similar $Q$.

The curves in Fig.1 have a monotonic dependence on temperature.
This suggests that the line separating compact and
extended states in the $(Q,T)$ plane is roughly straight,
starting from $Q=Q_c=q_0\sqrt{N}$ at infinite $T$.
To test this hypothesis, we considered the $Q$ and $N$ dependence
of the radius of gyration for chains of lengths $N=16$, 32, 64, 128.
To achieve good thermal averages, simulation were performed
at $T=0.4q_0^2/a$ and not at the lowest temperature in Fig.1.
 The dependence of $R_g^2$ on $Q$ is depicted in Fig.3; the vertical axis
is scaled so as to remove the trivial $N$--dependence for $Q=0$.
The charges on the horizontal axis are scaled to $Q/Q_c(N)$ for all
polymer lengths. Although $R_g^2$ is a monotonically increasing function
of $Q$, its variations are not gradual: for small $Q$
the radius barely depends on $Q$, while beyond a certain threshold
an extremely fast increase begins. This feature is further amplified
at low temperatures: the low charge polymers get smaller
while the charged ones increase in size.
Fig.3 strongly suggests that the transition
from compact to stretched configurations at low temperatures still
occurs for $Q\approx Q_c$. In fact, plotting $R_g^2(T=0.4q_0^2/a)/
R_g^2(T=\infty)$ as the function of $Q/N^{1/2}$, we find that the
curves become steeper with increasing $N$ and intersect at $Q/N^{1/2}
\approx 1.4q_0$, implying that for $Q>1.4Q_c$ the PAs at low $T$
are more stretched than self--avoiding walks.

To explain the above results, we start with the empirical observation
that the neutral PA compactifies to more or less a spherical shape.
This suggests an energy (or more correctly a quench averaged free energy)
of the neutral PA that grows with the number of monomers as
$E(Q=0)=-\epsilon_c N+4\pi R^2\gamma$,
where $\epsilon_c\propto q_0^2/a$ is the energy gain per particle,
and $R\approx aN^{1/3}$ is the radius, in the collapsed state.
A positive surface tension, $\gamma\approx q_0^2/a^3$, accounts
for the spherical shape of such PAs.
If we now uniformly add a charge $Q$ to each configuration, its energy
increases by $Q^2/R$. Therefore, $E(Q)\leq-\epsilon_cN+4\pi R^2\gamma+Q^2/R$;
the inequality indicating that the system can reduce its energy
by rearranging the charges. We would like to
find the rearrangements that minimize the energy. For the sake of
generality we shall consider $Q\propto q_0 N^\beta$, with
emphasis on the most relevant case of $\beta={1\over2}$.  To gain some
insight into the behavior of PAs we shall explore analogies to
charged drops. In the following paragraphs we briefly discuss the
shape of a charged {\it conducting} drop. This
analogy is most appropriate for an annealed version of the problem
in which the charges are free to move along the polymer chain.
We then go on to consider the shape of a charged {\it insulating} drop
(immobile charges), and finally conclude with applicability
of such arguments to quenched PAs.

With the charge on the surface of a freely suspended
{\it conducting} spherical drop of radius $R$, the non--extensive
contribution to the energy is
$E'(Q)=4\pi R^2\gamma+Q^2/(2R)$.
As $Q$ increases the shape of the drop may change. The surface energy
of an uncharged drop, $E'(0)=4\pi R^2\gamma$, sets the overall energy
scale of the problem, while the dimensionless parameter,
$$\alpha\equiv Q^2/(16\pi R^3\gamma) \equiv Q^2/Q^2_R\ ,$$
determines its shape. ($Q_R$ is the Rayleigh charge of the drop.)
Note that the above estimates
of $\gamma$ and $R$ for the model PA lead to $Q_R\approx Q_c$.
We shall initially consider only deformations in shape that maintain
the integrity of a single drop.
Such deformations of a charged conducting drop, and related problems,
have been considered by many authors\ref
\rgrig{See e.g., Grigor'ev A.I. and Shiryaeva S.O., Zh. Tekh. Fiz.
{\bf 61}, 19 (1991) [Sov. Phys. Tech. Phys. {\bf 36}, 258 (1991)];
and references therein.}.
It was noted by Rayleigh\ref
\rRay{Lord Rayleigh, Phil. Mag. {\bf 14}, 184 (1882).}\
that for $\alpha>1$ the charged sphere is locally unstable and must
thus become distorted. The problem was also studied by Taylor\ref
\rTaylor{Taylor G., Proc. R. Soc. London {\bf A280}, 383, (1964).}\
who considered distortions into prolate spheroids. For spheroids, the
energy and shape can be determined by minimizing a trial energy with
respect to the eccentricity, $e$. We can show\rKKtopub\ that for
$\alpha>0.899$ the energy of the spheroid is smaller than that of the sphere,
and the drop deforms into a strongly elongated shape. This qualitative
similarity to the experimental results\rExT, and to Fig.3 is encouraging.
While the drop strongly elongates for finite $\alpha$, it maintains a
finite aspect ratio. Gutin and Shakhnovich\ref
\rGSpp{Gutin A.M. and Shakhnovich E.I., preprint (1994).}\
recently reached a similar conclusion by assuming that the charged
PA distorts into a cylindrical shape with one long dimension $R_\parallel$,
and two short dimensions $R_\perp$ which are related by volume
conservation: Minimizing estimates of E(Q) for such shapes
leads to $R_\parallel\sim N^{(4\beta-1)/3}$, i.e.
no significant stretching for $\beta={1/2}$.
Another recent study by Dobrynin and Rubinstein\ref
\rDR{Dobrynin A.V. and Rubinstein M., preprint (1994).}\
relaxes the constant volume constraint and also reaches the conclusion
that there is an onset of stretching for $\beta={1\over2}$, although
a completely stretched state is reached only for $\beta={2/3}$,
when the Coulomb energy becomes extensive.

However, the spheroidal shape {\it is not} a local energy minimum\rTaylor\
as the strongly elongated spheroid is unstable
to a variety of perturbations\rgrig. Experimentally, for $\alpha>1$
the drop disintegrates into smaller ones.
Indeed if splitting of a continuum droplet is allowed, its
energy can be reduced to $E'(0)$. This is
achieved by splitting away from the original drop an infinite number of
infinitesimal droplets which contain all the charge
but have vanishing total electrostatic energy, surface area and
volume, and removing them to infinity.
Even if the particles are required to stay connected (to maintain the
connectivity of a chain), the system may be able to reduce its
energy by expelling its charge in the form of a finger.
Balancing the Coulomb energy ($Q^2/L$) of a finger of length $L$,
with a condensation energy $\epsilon_c L/a$, indicates $L\propto Q$.
Such fingers appear spontaneously if their cost (roughly
$Q\sqrt{\epsilon_c/a}$) is less than the Coulomb energy of the uniformly
charged sphere, $Q^2/R$, i.e. for $Q>R\sqrt{\epsilon_c/a}$.
Thus the typical {\it annealed} PA will have a protruding finger of
length $L\propto N^\beta$ for $\beta>1/3$. However, since the weight of
the finger is small, it will not effect the scaling of $R_g^2$ for
$\beta<5/9$. Such annealed PAs  have large spanning sizes without
appreciably greater $R_g$.

Now consider an opposite limit where the charge is uniformly distributed
along the polymer, mimicked by a droplet of charged {\it insulator}.
The sum of electrostatic and surface energies for a spherical drop is
$3Q^2/(5R)+4\pi R^2\gamma$. Unlike the conducting case, energy
cannot be lowered by expelling charge into small drops or fingers
(since distortions of charge and volume are now connected).
However, for $\alpha>0.293$ a drop can reduce its energy by
splitting into two equal drops separated by an in infinite distance.
Additional splittings occur for larger charges; the
optimal number of droplets increasing linearly with $\alpha$.
We can again constrain the overall object to remain singly connected
by attaching the droplets with narrow tubes of total length $L$ and
diameter $a$. As long as $L\ll aN$, most of the charge  remains in the
spheres. Equating the total electrostatic energy, $\propto Q^2/L$,
to the condensation cost, $\epsilon_c L/a$ (or a surface energy
of $\gamma a L$ in the continuum parlance), leads to $L\propto Q$.
Not surprisingly, the tubes in this case behave much the same as
the fingers for the conducting polymer. However, whereas the surface
tension in the conducting case results in one big central drop, for
the insulating case the droplets are separated as in a necklace.
The radius of gyration is now of the same order as the span of the
necklace, $L$.

How can we relate the above discussion to the quenched PA?
The immobile charges are likely to produce a necklace of compact beads
joined by narrow strings once the Rayleigh charge has been exceeded.
However, it may be possible for the random PA to
reduce its energy (compared to a homogeneous one) by taking
advantage of charge fluctuations along its sequence:
The strings will be constructed of segments with
a larger excess charge than average, while the beads are from
segments closer to neutral. Thus, $R_g\propto Q$ may underestimate
the size of the PA. The necklace picture may be hard to quantify
if the beads and strings have a broad distribution of sizes.
The following argument seems to suggest an upper bound to the
total length of the strings $L$ (and hence for $R_g$).
The string segments must lose at least a portion
of the condensation energy associated with compact beads. Since
this energy loss can only come from a reduction in the Coulomb
energy, we have $L<aQ^2/(\epsilon_cR)\propto N^{2\beta-1/3}$.
Thus while the quenched PA is certainly extended beyond the
Rayleigh instability, it seems difficult to justify
$R_g\propto N$.
However, we began the discussion
with rather robust assumptions on the condensation energy ($\propto N$)
and surface energy ($\propto N^{2/3}$), while the end--result involved
energies of order $N^{1/2}$. Consequently, an accurate estimate of the
stretching of the quenched PA requires a more detailed knowledge
its energy spectrum. A full understanding of the behavior of PAs is
yet to be attained.

We would like to thank B.I. Halperin for bringing the importance of
surface tension to our attention, and D. Ertas for helpful discussions.
This work was supported by the US--Israel BSF grant No. 92--00026,
by the NSF through No. DMR--87--19217  (at MIT's CMSE), DMR 91--15491
(at Harvard), and the PYI program (MK).

\vfill\eject\immediate\closeout\rfile
\centerline{{\bf References}}\bigskip
{\catcode`\@=11\escapechar=`  \input refs.tmp\vfill\eject}
\figures
\fig{1}{$R_g^2$ (in units of $a^2$)
as a function of $T$ (in units of $q_0^2/a$) for
several values of the excess charge $Q$ for a 64--monomer chain.
Each point is an average
over 10 quenches. Independent quenches are used at different temperatures.
The numbers near each curve indicate $Q/q_0$.}
\fig{2}{Histograms of the distribution of (ten) values of $R_g^2$
(measured for $N=64$ at $T=0.2q_0^2/a$) for several values of $Q/q_0$,
indicated near the histograms.}
\fig{3}{Scaled $R_g^2$ as a function of scaled charge $Q/q_0$ for
chain lengths $N=16$ (open triangles), 32 (full triangles),
64 (open circles), and 128 (full circles).}
\vfil
%\eject\null\vfil\centerline{\bf FIGURE 1}
%\eject\null\vfil\centerline{\bf FIGURE 2}
%\eject\null\vfil\centerline{\bf FIGURE 3}
\eject
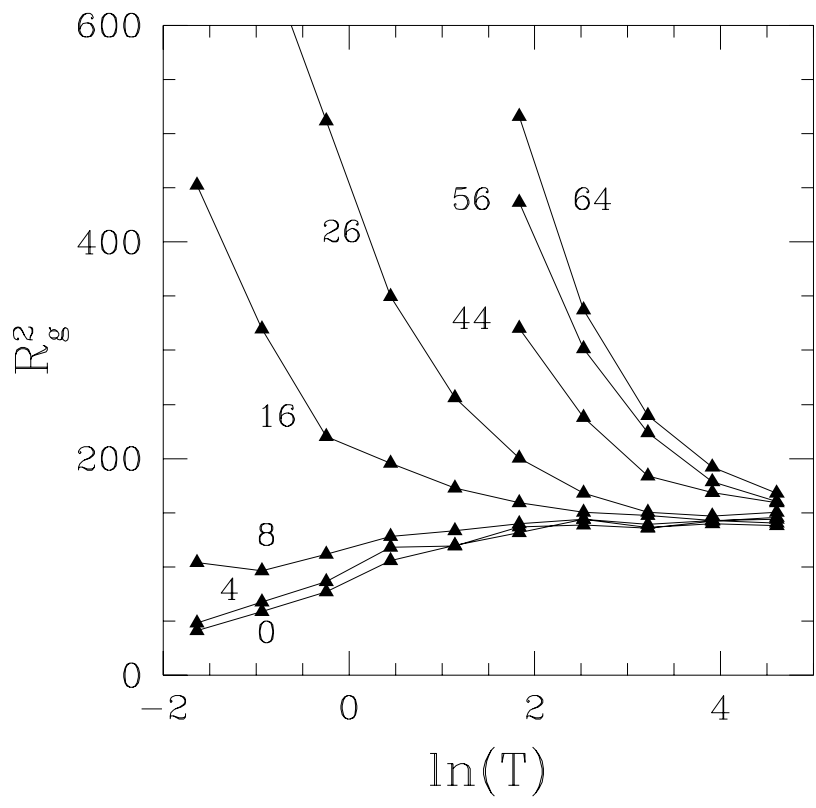
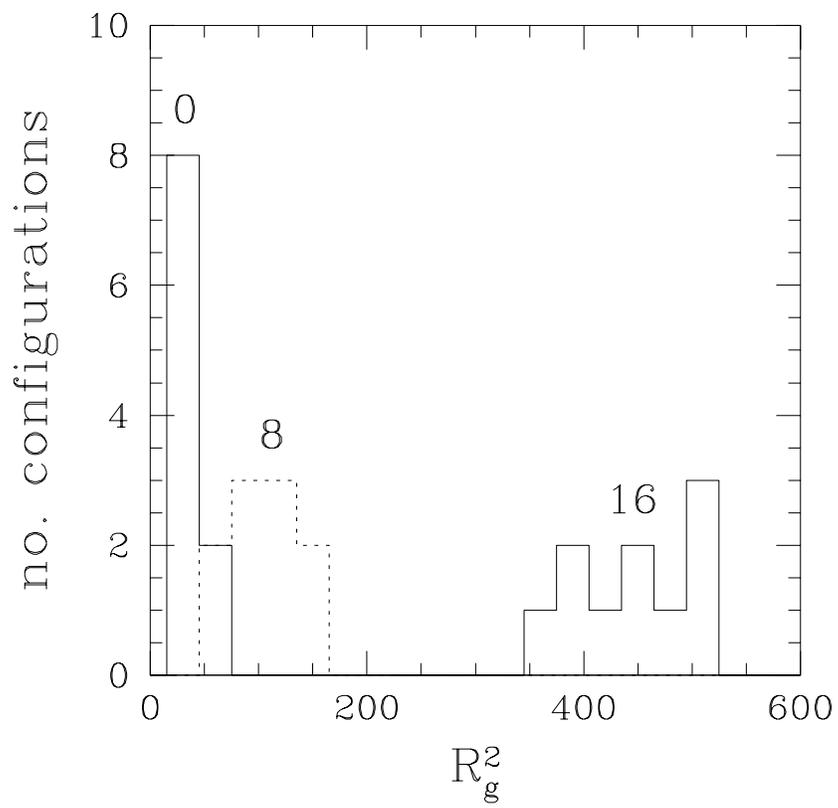
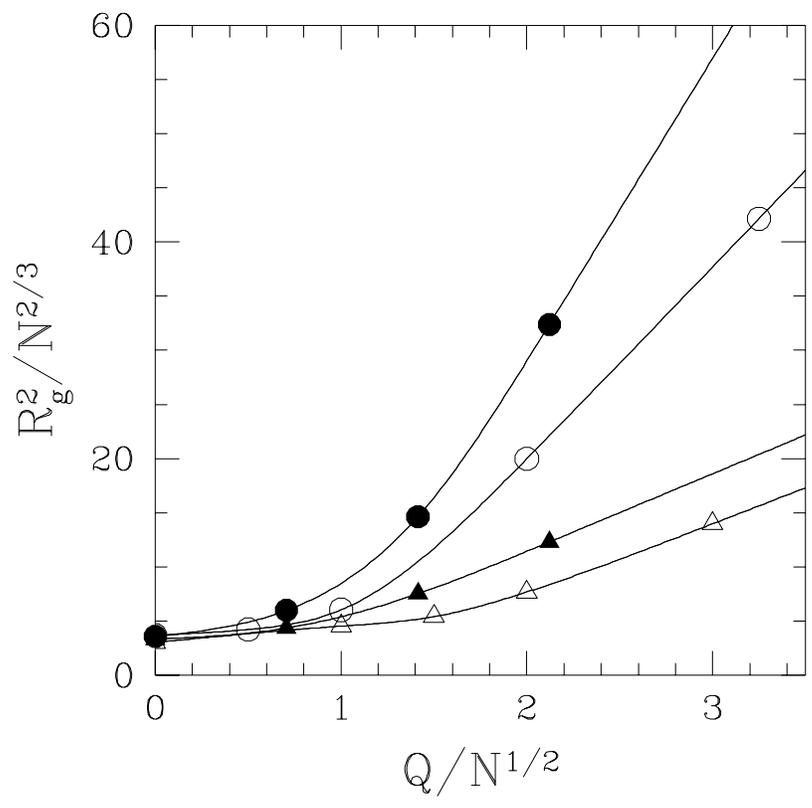
\bye